# Intuitive interaction flow：A Dual-Loop Human-Machine Collaboration Task Allocation Model and an experimental study

Jiang XU[1]，Qiyang MIAO[1]，Ziyuan HUANG[1]，Yilin LU[1,2]，Lingyun SUN[3]，Tianyang YU[1]，Jingru PEI[1]，Qichao ZHAO[4]

[1]School of Design and Innovation , Tongji University , Shanghai 200092;

[2]Jiangsu Thintron Industrial Design and Research Institute Ltd.；

[3]International Design Institute , Zhejiang University , Hangzhou 310058；

[4]Beijing Kingfar Technology Co.,ltd., Beijing 100083

**ABSTRACT:** This study investigates the issue of task allocation in Human-Machine Collaboration (HMC) within the context of Industry 4.0. By integrating philosophical insights and cognitive science, it clearly defines two typical modes of human behavior in human-machine interaction(HMI): skill-based intuitive behavior and knowledge-based intellectual behavior. Building on this, the concept of 'intuitive interaction flow' is innovatively introduced by combining human intuition with machine humanoid intelligence, leading to the construction of a dual-loop HMC task allocation model. Through comparative experiments measuring electroencephalogram (EEG) and electromyogram (EMG) activities, distinct physiological patterns associated with these behavior modes are identified, providing a preliminary foundation for future adaptive HMC frameworks. This work offers a pathway for developing intelligent HMC systems that effectively integrate human intuition and machine intelligence in Industry 4.0.

## 1 INTRODUCTION

The advent of Industry 4.0 has brought significant transformations in HMI, with the integration of automation and artificial intelligence (AI) technologies both substantially enhancing production efficiency and fundamentally altering traditional human-machine relationships [1-2]. Industry 4.0, also known as the Fourth Industrial Revolution, represents the convergence of advanced technologies designed to address significant societal and industrial challenges. It aims at achieving comprehensive digitalization and automation of the manufacturing sector through highly interconnected network systems and intelligent technologies [3-4]. Intelligent machines equipped with multiple sensors can acquire critical operational parameters and external environmental data in real-time, enabling them to optimize response strategies and achieve operational precision and efficiency through data analysis [5-6]. Consequently, many tedious, repetitive, and potentially hazardous tasks previously performed by humans are now taken over by highly automated intelligent machines, demonstrating the potential for full automation in specific scenarios [7]. However, given the complexity and unpredictability of operational environments, coupled with current technological limitations, achieving full automation of tasks solely through machines presents significant challenges. Human participation remains essential.

In response to these challenges, HMC systems have emerged as a central focus of research. This paradigm shift from standalone automation to HMC underscores the complementary nature of human and machine intelligence, aiming to integrate heterogeneous intelligences to jointly accomplish complex tasks. The integration of heterogeneous intelligence poses a significant challenge in managing complexity, and current task allocation mechanisms often fail to fully utilize human resources and skills. Particularly as intelligent machines take on more automated tasks, humans are gradually shifting toward supervisory and management roles. However, intelligent machines have their limitations in addressing complex, nonlinear tasks, and the intuition and nonlinear perception abilities demonstrated by humans remain irreplaceable. At the same time, merely being in a supervisory or takeover role may lead to the degradation and loss of essential human coping skills, resulting in failures in human-machine collaborative task responses. Therefore, the core value of HMC lies in maximizing the utilization of both human and machine resources, effectively balancing the respective advantages of human intelligence and machine intelligence to achieve optimal task allocation.

Henri Bergson and Hubert Dreyfus have incorporated issues such as human intelligence, cognition, and skills into the philosophical domain [10-12]. Their discussions on concepts like intuition offer profound insights for re-examining the relationship between human-machine interaction and the unique advantages of human intelligence. Therefore, this study aims to introduce a philosophical perspective, integrating philosophical insights with practical phenomena. We explore the fundamental differences in intelligent performance between humans and machines, clearly defining two typical human behavior modes in human-machine interaction: skill-based intuitive behavior and knowledge-based cognitive behavior. This study focuses on combining the human advantage of intuition with the machine's intelligent capabilities, aiming to clarify a possible fundamental mechanism of task allocation in HMC.

Furthermore, HMC relies on bidirectional interaction. Both humans and machines must understand each other's intentions, provide mutual feedback, and ultimately adapt to one another. Therefore, constructing an HMC framework requires modeling human cognitive, perceptual, motor, and affective factors to help machines recognize and adapt to changes in human conditions during task execution. To explore the specific physiological characteristics underlying typical behavior patterns in actual human-machine interaction tasks, this study conducts comparative experiments. By measuring and comparing the differences in EEG and EMG activities between experts and novices while performing specific tasks, the study aims to reveal the neurophysiological patterns associated with these two behavior modes, providing empirical evidence for recognizing the human state during HMC.

## 2 RELATED WORK
### 2.1 Intuition theories
#### 2.1.1 Characteristics of intuition: Unconsciousness and continuity

Human beings possess a nonlinear, intuitive capacity to respond, which aids in making rapid decisions and taking actions in complex, dynamic environments. Previous studies have attributed this cognitive and action ability, which transcends rationality, to human intuition[13-14].

The concept of intuition, while historically significant, lacks a mainstream or gold standard in its study[13]. Various paradigms and theoretical models prevail across different disciplines. In psychology, intuition is characterized by its ability to handle uncertainty, process coherence, and unconscious nature. It evolves as a continuous and dynamic cognitive process, developed through extensive practice in specific fields, which culminates in deep expertise. This expertise is crucial for rapid decision-making, especially in navigating complex and uncertain environments. Piaget describes intuition as an irrational mental function that acquires knowledge through unconscious processes, evolving into an ability for rapid judgment that operates independently of conscious thought [15]. Claxton et al. emphasize that intuitive tendencies result from an experience-based unconscious process, which subsequently acts as a strong signal initiating action, enabling rapid response based on one's intuitive perceptions [16]. Bowers et al. further elucidate the role of intuition in guiding human actions, positing that intuition is a coherent structure, a preliminary perception of coherence, which then guides thoughts and actions towards the inherent nature of coherence [17]. In other words, the intuitive process, through sharp extraction and organizational grasp of information in complex situations, displays coherence, enabling a swift and flexible response to complex scenarios.

Philosophical discussions often highlight the critical role that human perception and cognition play in the concept of intuition. Henri Bergson identified two fundamental ways of understanding the world: 'intellect' and 'intuition.' He describes 'intellect' as a method skilled in handling the static material world through scientific and logical thought, systematically breaking down complex entities into analyzable parts [18]. Intuition, in contrast, is a deeper cognitive process that directly accesses the essence and inner vitality of things. According to Bergson, 'intellect' operates like 'beads on a string,' using discrete, isolated elements. In stark contrast, 'intuition' unfolds as a seamless, continuous process that skillfully 'bridges the gaps' between concepts, fostering a fluid and ongoing stream of thought [19]. Whereas 'intellect' tends to render the world static and immutable, 'intuition' captures its fluid and ephemeral nature [20]. Therefore, intuition acts as a fleeting 'impulse' that transcends analytical understanding, providing individuals with a direct gateway to the essence of things [19]. This mode engages individuals in an immersive, fluid interaction with the world, thereby connecting our immediate experiences to the 'unique' and 'inexpressible' qualities of objects. Deleuze introduces the concept of 'attention following' to frame intuition within the realm of action [21]. Through 'attention following,' individuals navigate and interact with the real-time complexities of their environments, making intuitive leaps that are deeply informed by their engagement with the present. This engagement is not passive, but a dynamic, embodied practice. It is through such an intuitive process—a continuous interplay of perception and action with the world—that individuals build a genuine understanding and enter into a more continuous and unified world of experience. In this experiential realm, actions are not mechanical reactions to the external world but are part of an internally generated process characterized by inherent continuity [22]. Dreyfus similarly defines intuition as a rapid, fluid, and engaged mode of perception and action [23]. He views intuition as the fusion of ingrained bodily behavior and perception, grounded in the individual's extensive experience and sensitivity to the environment [23]. This integration typically leads to decision-making without conscious involvement in the details, allowing the individual to act swiftly, efficiently, and with coherence and adaptability.

In summary, despite variations among theories of intuition, there is a shared consensus on its unconscious, continuous, and fluid nature. Intuition is widely recognized as an advanced cognitive ability rooted in extensive experience, in which repeated practice and engagement form the foundation of intuitive behavior.

**2.1.2 Intuitive Coping and Embodied Skill Acquisition**

In the field of AI, human intuition, noted for its speed and accuracy, has attracted considerable attention. Simon and other AI scholars regard intuition as an efficient information-processing mechanism that enables rapid decision-making without explicit logical reasoning. Simon describes intuition as a process wherein individuals leverage prior knowledge and experience to swiftly recall and identify appropriate solutions to problems [24]. From this perspective, he proposes that machines can simulate human intuition through basic pattern recognition mechanisms. By encoding human expertise into algorithms, machines can acquire high-level knowledge, enabling them to make rapid judgments in complex scenarios [14].

This perspective differs significantly from the discussions of intuition in the field of phenomenology. From Dreyfus's point of view, intuition, as an advanced ability developed through long-term skill acquisition and practice, fundamentally distinguishes humans from intelligent machines. Dreyfus directly associates intuition with skill, viewing it as representative of the highest level of expert practice within a given field [23]. "Human experts, after years of experience, are able to respond intuitively to situations in a way that defies logic [23]." Dreyfus's skill acquisition model outlines how individuals progress through distinct stages of skill development, from novice to expert. The behavior of novices, characterized by a slow and deliberate adherence to rules, closely resembles the cognitive approach Bergson defines as "intellect". Novices rely on analytical thinking and careful deliberation to perform tasks, following predefined guidelines. In contrast, the behavior of experts is immediate and intuitive, aligning with what Bergson refers to as "intuition" — a response rooted in a deep understanding of context, without the need for conscious analysis. As practice and skill development advance, the rules that novices initially depend on are not rigidly internalized or memorized [11]. Rather, they are embodied in a non-theoretical manner through the body's direct engagement with the task. This shift enables intuitive responses to gradually replace rational, rule-based reactions, facilitating expert-level agility and responsiveness. At this stage, the intrusion of semantics or rules can actually disrupt the expert's optimal grasp of complex situations. Without the need for conscious awareness or thought, the subject exhibits a deep immersion in experience, along with heightened sensitivity and adaptability to the environment. Intuition is rooted in the subject's well-honed embodied skills, allowing them to fluidly respond to tasks through ingrained habitual bodily patterns, entering a state of continuous, seamless flow of direct bodily experience [25].

These philosophical concepts are also supported in the field of cognitive psychology. The most widely accepted view in contemporary cognitive psychology is that intuition is inherently related to pre-established knowledge formed through implicit memory and learning [26-27]. Additionally, there is a dual-system thinking model hypothesis. The dual-system theory, also known as the dual-process theory, suggests that human cognition operates through two distinct systems or modes of thinking: System 1 and System 2 [28]. System 1 corresponds to intuition and intuitive thinking, which is fast, automatic, and often unconscious. This intuitive thinking is quick and effortless, relying on heuristics or mental shortcuts, and operates based on tacit knowledge, often linked to experience-based learning and gut feelings. It excels at handling familiar situations, making rapid judgments, and responding to immediate demands without the need for deliberate thought, and is associated with procedural and implicit learning. System 2 corresponds to intellect and rational thinking, which requires effortful and logical reasoning, involving explicit knowledge. It is engaged in tasks that demand careful consideration, such as solving complex problems or coping with unfamiliar tasks, and is associated with explicit and declarative learning. These two systems complement each other, working together to enable effective decision-making across a range of situations. However, through repeated practice and learning, processes initially handled by System 2 can become more automatic and shift into System 1 over time, allowing them to be executed more quickly and effortlessly. Intuitive thinking is considered the culmination of implicit learning, through which individuals acquire the knowledge necessary to make intuitive judgments about specific matters [29]. The process of acquiring intuition involves a transition from explicit knowledge to tacit knowledge after extensive skills and experience are accumulated through long-term experiential practice and perception, reflecting a shift from System 2 to System 1. At the same time, there is a shift in how we engage with the relevant domain, moving from an initial representational approach to a fully non-representational, purely embodied, and enactive mode of engagement. In summary, the generation and accumulation of tacit knowledge is an unconscious and gradual process, intricately intertwined with the subject's rich personal experiences. This makes tacit knowledge deeply embedded in personal experience, skills, and insights, rendering it difficult to articulate and distinct from implicit knowledge.

The core logic of scientific methods is largely based on quantitative abstraction through symbolic representation. By translating complex realities into quantifiable terms using symbols and equations, science seeks to uncover underlying patterns, relationships, and laws. The design logic of AI and machines is fundamentally rooted in these scientific methods. AI's simulation of human intuition similarly relies on structured, formalized, and quantifiable explicit knowledge, using discrete units of data to establish fixed logical and formal rules. This approach allows algorithms and models to efficiently analyze and interpret large datasets, extracting valuable information and patterns to make decisions or carry out specific tasks. However, the essence of this method is spatial and sequential, fragmenting the complexity and continuity of the real world into static, geometric points and mathematical instances, which inherently falls short in fully capturing the fluid and continuous nature of human intuition. While machines can learn from partial representations and symbolizations of human tacit knowledge, they remain constrained by the limitations of explicit knowledge and formal rules. Tacit knowledge, by contrast, is inherently situated and contingent on local conditions and understandings, always executed within the cognition and actions of the subject. This type of knowledge is not object-like and cannot be fully transferred or codified. Consequently, no form of simulation based on implicit knowledge can fully account for the nuances of specific, situated intuition. In other words, the mechanisms through which human tacit knowledge operates are difficult for machines to replicate. The complexity of real-world situations often exceeds the capabilities of purely formal algorithms, making the continuity and fluidity of human intuition irreplaceable and highlighting the inherent limitations of machines in fully simulating human cognitive processes.

## 2.2 HMC

### 2.2.1Task Allocation Model

In early research on HMC, the Fitts List adopted a static task allocation model that assigned basic functions to the most suitable entities based on their execution efficiency. This model provided a guiding framework to determine whether a specific function should be executed by a human, a machine, or a combination of both [30]. However, with continuous technological advancements, humans' traditional advantages in certain tasks have been replaced by intelligent machines. Moreover, this approach has been widely criticized for its lack of flexibility, as it cannot adapt to the dynamic changes in environmental and task requirements. As a result, it has gradually been replaced by Dynamic Function Allocation (DFA) models [31].

The early DFA model assumes that task allocation should dynamically adjust based on the changing cognitive state of the human operator. The system intervenes when human information processing limitations arise, helping to overcome bottlenecks and meet operational requirements [32]. As AI technologies continue to advance, intelligent machines are gradually demonstrating the potential to complete tasks, achieve goals, and interact with their environment independently, without human intervention. In the DFA model, machines are taking on an increasing number of automated tasks, while the human role is shifting from direct handling and operation to supervision, management, and intervention.

Nevertheless, in highly complex or uncertain tasks, even well-designed intelligent systems can deviate or fail when operating independently. While intelligent machines excel at processing structured information and solving symbolic problems, particularly in discrete tasks, they often lack the ability to break rules or respond flexibly. In contrast, humans possess a nonlinear, intuitive capability that offers an irreplaceable advantage in complex, dynamic, and unstructured tasks. Therefore, task allocation in HMC must integrate human nonlinear intuitive capabilities into the feedback loops of intelligent systems. This integration can ensure a close coupling of human intuitive responses to complex and uncertain problems with machine intelligence systems.

In complex and dynamic situations, when automation fails, human operators must respond quickly. However, extended periods in a regulatory role often reduce workload, causing distractions. They are required to transition almost instantly from a low-workload, low-pressure state to handling high-workload, high-pressure tasks, which can impair performance and lead to failures in the overall collaborative system. Furthermore, when machines perform most routine tasks, human skills may degrade, resulting in reduced practical experience and diminished ability to cope with complex, uncertain tasks. Bainbridge's "Automation Paradox" highlights the core issue of human-automation coupling failure: increased machine automation reduces opportunities for humans to maintain critical skills during tasks. This over-reliance on automated systems neglects the unique value humans contribute to decision-making and problem-solving, gradually eroding their ability to effectively fulfill their roles within the HMC system [33]. Therefore, task allocation in HMC must address how to keep humans in an optimal state to leverage their critical advantages in complex situations.

In summary, as HMC becomes central to intelligent system operations, effective collaboration demands clear task allocation. In complex and uncertain tasks, the unique value of human nonlinear intuitive response capabilities should not be overlooked. Additionally, task allocation between humans and machines must fully consider human operational needs, ensuring active human participation to maintain responsiveness. Thus, the role of humans extends beyond merely monitoring and managing intelligent machines. It is essential to define how to achieve efficient HMC through rational task allocation, maximizing the strengths of both humans and machines.

### 2.2.2 HMC Adaptive Framework

Continuous and effective HMC relies on close coordination and feedback. Thus, beyond establishing a task allocation mechanism, successful HMC requires the clear definition of communication and interaction methods to establish a closed-loop feedback system. Cassenti et al. identify four types of interaction between humans and intelligent systems: user-initiated action, concurrent performance, physiological variables, and cognitive modeling [34].

User-initiated actions occur when the user, through self-awareness and assessment of task-related difficulties, actively triggers system support via commands, thus initiating collaboration with the intelligent system. In the early stages, user-initiated actions were typically conveyed through physical or graphical interfaces [35]. With advancements in machine intelligence, human-machine systems now incorporate multimodal interaction methods such as natural language understanding [36] and gesture interaction [36], offering enhanced communication tools.

Concurrent performance is evaluated by monitoring task performance against a predefined threshold. The system continuously tracks performance metrics, calculating when performance falls below the set threshold within a specific time interval. When the threshold is breached, the system activates auxiliary tools. Parasuraman et al. [37] use performance measures, including accuracy and reaction time, as inputs for cooperative systems. The system then triggers automation based on user performance indicators, such as task completion or response speed. Feigh et al. [38] further propose an adaptive collaboration model based on performance thresholds, which is especially effective in tasks and domains with clear constraints and standards.

Physiological variables involve the system continuously monitoring users' cognitive states through real-time physiological indicators such as heart rate, EMG, and EEG [39]. Increasingly, studies are using large-scale physiological data, collected during task execution and analyzed with machine learning techniques [36], to rapidly distinguish different user states during HMC. This allows the system to recognize human behavioral states and provide adaptive assistance accordingly.

Cognitive modeling simulates human thought processes to predict human states and behaviors, thereby offering appropriate assistance. Common models include LICAI, EPIC, and ACT-R [40].

In these methods, user-initiated actions are constrained by human self-awareness and subjective factors. When users fail to accurately assess the situation or their own efficacy, the intelligent system may not be effectively activated for collaboration. Although performance thresholds offer an objective means for assessing the current state, this approach assumes that task performance can be evaluated solely through fixed indicators, failing to account for the dynamic changes of complex real-world scenarios. Additionally, performance calculations often lag, hindering the system's ability to reflect real-time fluctuations in human performance. In contrast, physiological variables provide a more dynamic and flexible approach. As continuously measurable objective indicators, they offer direct insight into human states. However, human physiological states are linked to various cognitive processes, and traditional methods often rely solely on cognitive load or attention levels as triggers for automation, insufficiently capturing the full complexity of human cognition. However, cognitive modeling simplifies human cognitive processes, viewing humans as limited computational entities, and overlooking the dynamic nature of human perception and behavior in real-world contexts.

Therefore, establishing an ideal human-machine collaborative feedback loop requires a deeper understanding of human cognitive processes, which can lead to the development of optimized monitoring and feedback systems. One promising approach is to collect and analyze behavioral data through cognitive models and empirical research, identifying key indicators and behavioral variables for adaptive collaboration. By continuously perceiving human states in real time and pinpointing critical indicators, machines can adaptively adjust to human conditions, providing appropriate support. This creates dynamic, continuous, bidirectional interaction between humans and machines, ultimately achieving tight integration and efficient collaboration.

## 3 DUAL-LOOP HMC TASK ALLOCATION MODEL

### 3.1 Intuitive interaction flow

The design of intelligent machine systems is based on clearly defined rules and algorithms, following an objective and discrete mode of knowledge processing, making it challenging to simulate the deep and non-linear characteristics of human intuition. In the context of human-machine interaction, the states of intuition and intellect exhibited by experts and novices represent two typical conditions of interaction between humans and intelligent machines. The novice stage corresponds to a knowledge-based interaction mode, where humans use representable conceptual knowledge to analyze situational information based on task requirements and take corresponding actions. As Heidegger describes, in the 'present-at-hand' state, tools are in hand but are not yet used proficiently, thus failing to achieve seamless integration with their user [41]. This mode signifies a disruption of natural, fluid engagement typical of tool use. Here, tools are perceived outside their everyday use context, regarded purely as objects that are observable and analyzable from an external standpoint. This suggests that, at this stage, the subject cannot fully immerse in the direct execution of tasks, as they must divert their attention to the contemplation and evaluation of machine operation. The expert stage corresponds to a skill-based interaction mode, where skilled experts no longer rely on rules but are fully immersed in the world of the skill, intuitively responding to situational demands with skillful coping, possessing a deep sense of engagement with the situation. As in the practice of driving behavior, "The expert driver becomes one with his car, and he experiences himself simply as driving, rather than as driving a car, just as, at other times, he certainly experiences himself as walking [23]." At this stage, the machine is experienced as 'transparent,' almost as if it 'disappears' from the subject's horizon. As Heidegger described in the state of 'readiness-to-hand,' the tools integrate into the user's bodily space, assisting them in smoothly completing a series of object-oriented activities. The human subject has entered a state of complete flow and continuous intuition, with human-machine interaction transcending traditional subject-object separation, forming a deeply integrated symbiotic relationship.

Human intuition fosters the ideal state of human-machine fusion, in which the flow and continuity of intuition offer a new perspective for tuning and collaboration. On the one hand, guided by intuition, the human subject requires almost no conscious effort in highly automated skillful actions, resulting in a smooth and effortless state of action that maintains sustained efficiency in task execution. On the other hand, in this intuitive state, the human subject and the machine are highly integrated; the machine system acts as a direct extension of human perception and action by offering appropriate assistance. The human-machine system responds to external task demands with optimal efficiency, ensuring strong performance. Accordingly, the study defines intuitive interaction flow as a highly integrated and seamless working state formed through the dynamic interaction between human intuitive behavior and machine intelligence within a HMC system. In this state, the human operator is fully immersed in the task, with their perception and actions exhibiting non-reflective continuity, allowing the operator to act effortlessly. The machine continuously monitors the operator's state, making adaptive adjustments and providing appropriate assistance to ensure that the task's difficulty aligns with the operator's skill level, thereby promoting and sustaining the operator's engagement in this continuous state. Through such dynamic adjustment and adaptation, human and machine maintain a continuous, fluid bidirectional interaction, ultimately achieving coordinated synchronization and organic collaboration, ensuring optimal cooperation during task execution.

To further enhance the flow in HMC, it is essential to identify and leverage the flow state of the human operator during interaction. This enables the machine to make adaptive adjustments and optimizations based on the operator's needs and condition, ensuring that the operator remains in or closely approaches a state of flow while also improving the system's efficiency in task management. Specifically, when machines detect that human subjects are engaged in skill-based intuitive stages, the subjects' cognitive and behavioral states exhibit highly continuous, non-reflective, and intuitive characteristics. At this point, explicit rules or directives can negatively impact the subjects' actions, as excessive machine intervention may disrupt this continuity. Therefore, machines should provide timely and precise feedback only when necessary, minimizing unnecessary interference to maintain the operator's immersed state and facilitate efficient collaboration. Conversely, when subjects are engaged in knowledge-based intellectual stages, the complexity of the task environment or a lack of proficiency in required skills often leads to increased cognitive load and decreased operational efficiency. As Martin Heidegger describes, in the 'present-at-hand' state, tools are in hand but cannot yet be used proficiently, failing to achieve a seamless integration with their user [41]. This mode signifies a disruption to the natural, fluid engagement typical of tool use. Here, tools are perceived outside their everyday use context, regarded purely as objects that are observable and analyzable from an external standpoint. This suggests that, at this stage, the subject is unable to fully immerse in the direct execution of tasks, as they must divert their attention to the contemplation and evaluation of machine operation. While intelligent machine systems can fully utilize their computational and reasoning capabilities by integrating advanced technologies and AI algorithms, these systems can proactively sense complex environments, accurately recognize human behavioral states, and provide timely or even preemptive support or intervention. This assistance aids humans in making complex judgments and decisions, thereby helping them maintain behavioral continuity and enhancing their sense of immersion. In other words, when addressing knowledge-based human behaviors, machines are expected to offer greater decision support, assisting users in analyzing situations, understanding rules, and subsequently making more informed decisions.

In summary, effective HMC can be understood as a bidirectional matching and tuning process, where the intuitive interaction flow represents a continuous and organic state of engagement between humans and machines. Ideally, machine design should integrate the intuitive strengths of humans with the reasoning advantages of machines to achieve an optimal balance in task allocation. Building on this perspective, we introduce a preliminary framework for HMC through the lens of intuitive interaction flow (as shown in Figure 1), with the aim of maintaining or restoring continuity in the interaction process.

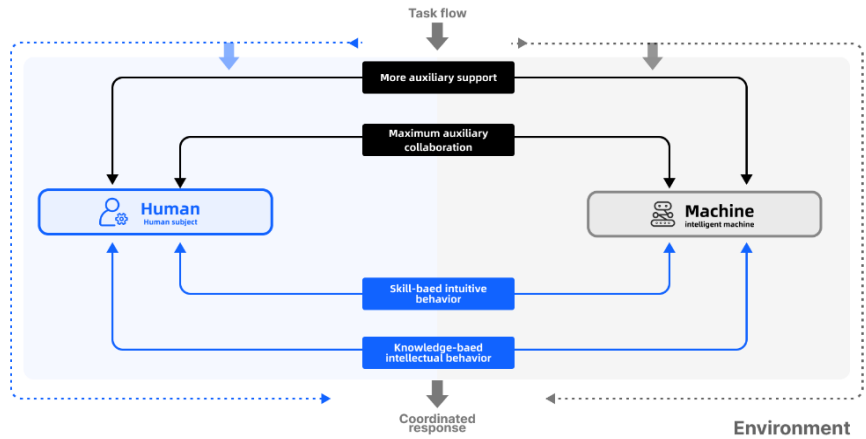

Figure 1 HMC model driven by intuitive interaction flow

### 3.2 HMC task allocation adaptive framework

In actual task scenarios, human behavior and cognitive states are continuously evolving. Therefore, this study adopts a human cognitive architecture model to identify the neural correlations underlying human intuitive and intellectual states, thereby aiding intelligent systems in recognizing and supporting human actions to ensure the adaptability and efficiency of collaborative systems. The LIDA (Learning Intelligent Distribution Agent) model, proposed by Franklin et al., aims to enhance the processing of complex tasks by AI systems through the simulation of human cognitive activities [41-43]. This model integrates various cognitive science theories, including Global Workspace Theory (GWT), grounded cognition, and long-term working memory, to provide a comprehensive and holistic analysis of cognitive processes. By incorporating embodied cognition theory, the LIDA model simulates the dynamic coupling of human perception and action, further analyzing the mechanisms underlying the formation of human skills. It conceptualizes human behavior as the activation and execution of a series of behavioral schemas, which become increasingly automated through repeated practice and accumulated experience, eventually internalized as skills that enable proficient responses in complex tasks. Thus, the LIDA model provides a theoretical foundation for explaining the various cognitive and behavioral states during human interaction with intelligent systems.

The study simplifies and adapts the LIDA model (as shown in Figure 2), categorizing human cognitive and behavioral loops into the skill-based loop and the knowledge-based loop. The skill-based loop involves highly automated behaviors that develop through long-term practice. External environments are perceived and stored in sensory memory and are associated with both the current body schema and the habitual body schema through the dorsal stream. The current body schema provides information about the human subject's present state (e.g., body position, limb angles), while the habitual body schema contains learned skills stored in the perceptual associative memory module [42]. When sensory memory appropriately matches both the current body schema and the habitual body schema, it indicates that the human subject possesses the relevant skills and is in a state conducive to executing behaviors within the skill-based loop. At this point, the situational model directly accesses procedural memory to automatically execute motor plans without the need for consciousness allocation through the global workspace, allowing for the automatic execution of highly automated behaviors [42]. Simultaneously, sensory memory directly influences the motor planning module via the dorsal stream, achieving a direct coupling between perception and action, which allows real-time fine-tuning to maintain operational status. However, when sensory memory fails to fully match with the habitual body schema, the subject continues to choose actions based on the existing skill-based loop but also requires consciousness allocation through the global workspace to update the action selection mode, thereby reinforcing the perceptual associative memory and facilitating further updates to the habitual body schema, or the skills. The knowledge-based loop involves scenarios requiring conscious decision-making and deliberation, where the human subject lacks the necessary skills, resulting in a complete mismatch between sensory memory and the habitual body schema [42]. In this case, behavior is entirely based on action mode selection mediated by the global workspace. As the behavior is repeated, the corresponding action modes also influence the perceptual associative memory, gradually forming behaviors within the skill-based loop.

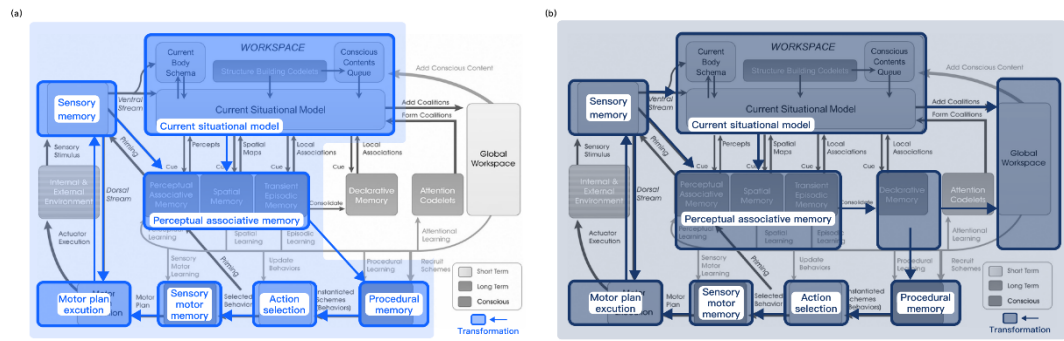

Figure 2 (a) Intuitive loop transformation process based on LIDA model, (b) Intellect loop transformation process based on LIDA model

Corresponding to the skill-based loop and the knowledge-based loop, the study further constructs a dual-loop cognitive-behavioral model（as shown in Figure 3）based on the former HMC model, centered around two core loops: the intuitive loop and the intellectual loop. In the intuitive loop, at the onset of a task, both the human subject and the machine simultaneously respond to the inflow of tasks from the external environment. The human subject automatically matches sensory memory with the current body schema and habitual body schema based on the present state information. If a match is successful, the subject quickly enters the skill-based behavior execution phase. Correspondingly, once the intelligent system identifies the human's skill-based behavior pattern, it provides minimal adaptive assistance to avoid disrupting the user's intuitive behavior. In the intellectual loop, if sensory memory fails to fully match the habitual body schema, the subject exhibits a knowledge-based behavior pattern, and the intelligent system provides corresponding decision support. This process iterates continuously, with humans and intelligent systems optimizing task execution strategies and adjusting action plans based on external environmental changes and dynamic feedback between them.

Based on the dual-loop cognitive-behavioral model, we summarize the physiological characteristics corresponding to the two states. The skill-based loop relies on the subject's internalized knowledge and experience, involving minimal conscious judgment and logical analysis. In this state, cognitive load is relatively low, and bodily experience directly guides behavior, characterized by a high degree of perception-action coupling. In contrast, the knowledge-based loop typically requires significant logical thinking, analysis, and decision-making, leading the subject to operate under a higher cognitive load. In this state, the subject has not yet developed an automatic perception-action pattern, resulting in a low degree of perception-action coupling.

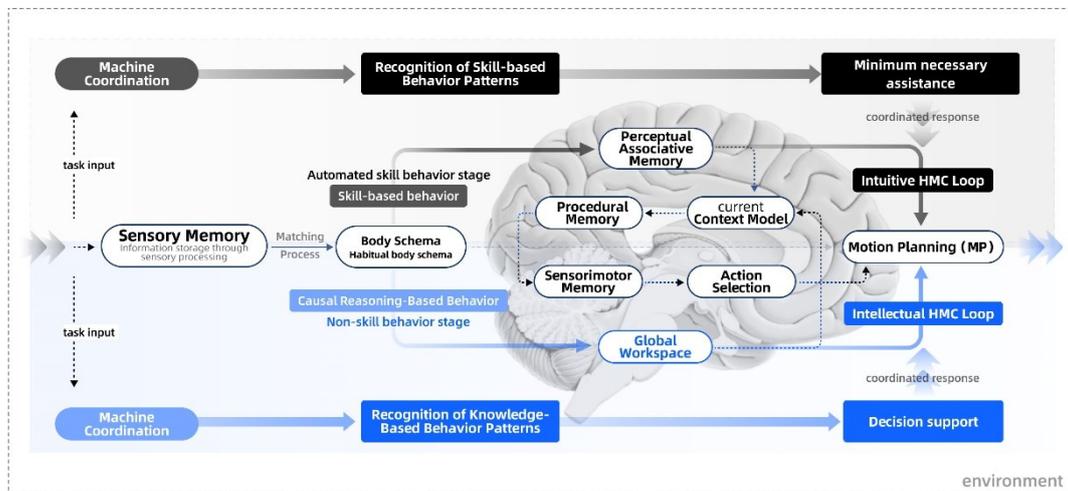

Figure 3 A dual-loop cognitive-behavioral model

Based on the aforementioned model, it can be inferred that human performance during task execution exhibits two typical states—intuition and intellect—which can be distinguished by cognitive load and the characteristics of perception-action coupling. These states can be represented through physiological signals such as EEG and EMG. In the context of HMC, real-time monitoring of these physiological indicators is conducted. Data preprocessing methods, including Independent Component Analysis (ICA), band-pass filtering, high-pass filtering, and low-pass filtering, are employed to eliminate artifacts from sources such as heartbeat and muscle activity [44,45]. Subsequently, feature extraction techniques, such as Fast Fourier Transform (FFT) and wavelet transform, are applied to analyze the signals. Based on the results of feature extraction, machine learning algorithms, including Support Vector Machines (SVM), decision trees, and random forests, are utilized for classification [46,47]. This process effectively distinguishes between the intuitive and intellectual states of the subject during task execution. Through these classification results, intelligent systems can dynamically adjust their assistance strategies to better meet the actual needs of the subject (as shown in Figure 4).

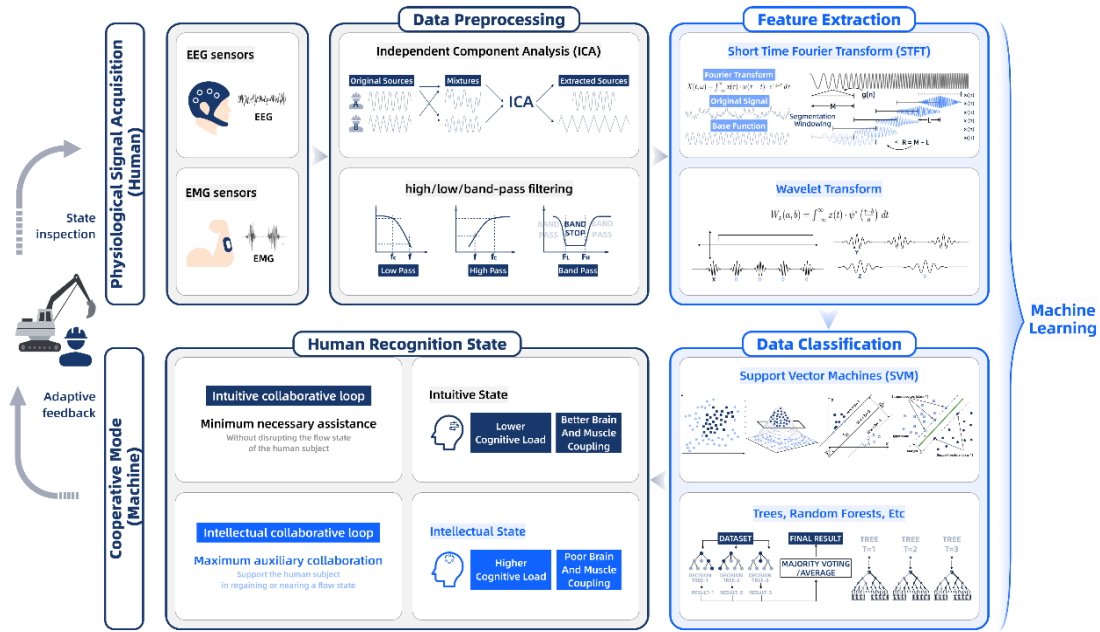

Figure 4 Dual-Loop HMC Adaptive Framework

## 4 MATERIALS AND METHODS

To explore the specific physiological characteristics underlying typical behavior patterns in human-machine interaction tasks, this research designs an experiment. By measuring and comparing the differences in EEG and EMG activities between expert and novice participants during the execution of specific tasks, this research aims to preliminarily reveal the neurophysiological patterns related to these two behavior modes, providing empirical basis for recognizing the state of human subjects in HMC. We hypothesize that the distinction between intuitive and intellectual states in humans can be discerned through physiological measures, including EEG and EMG indicators. Experts are expected to predominantly demonstrate intuitive states during task execution, characterized by lower cognitive load and higher synchronization between perception and behavior, while novices are expected to primarily exhibit intellectual states, characterized by higher cognitive load and lower synchronization between perception and behavior.

### 4.1 Experimental protocol

Excavators are indispensable pieces of machinery in the construction machinery industry, operating in varied environments with complex tasks. Traditional excavator operation primarily relies on joysticks and other physical control elements, such as levers and switches. These operations require a high degree of coordination and spatial imagination, and it typically takes operators many years to master the necessary skills [48,49]. Due to the dynamic, complex, and uncertain nature of construction sites, most excavator tasks in real-world scenarios still require execution by highly skilled expert operators [50,51]. Although excavator operators generally undergo similar training, their ability to handle tasks can vary significantly due to individual factors such as age, expertise, and practical experience. Previous studies have shown that it generally takes 5 to 10 years of operational experience for an excavator operator to reach an expert level of proficiency [50]. As the demand for construction projects continues to grow and the industry faces an aging workforce, there is a notable shortage of skilled operators who can achieve high levels of performance in excavation tasks. Given the trends towards automation and intelligent systems, future excavator HMC systems must incorporate the expertise of seasoned operators while also providing robust support for novices who lack professional knowledge and practical experience, thereby enabling them to perform tasks more effectively.

The routine operational tasks of excavators include digging, loading, crushing, trench-finding, and backfilling. The trench-finding task, used to identify underground pipelines, cables, or other infrastructure to prevent damage during large-scale excavation or construction, requires the operator to exercise high levels of judgment and operational skill. Digging and trench-finding were selected as excavation tasks to elicit differentiated behavioral modes between expert and novice groups under different levels of task complexity. In the experimental site designed to simulate the daily working environment of excavators, the participants successively completed the tasks of digging and trench-finding according to the specified requirements （as shown in Figure 5）. In the trenching task, a 1.5-meter by 2-meter area is designated in advance, and participants are required to perform trenching actions within this area. The trenching action consists of six cycles, with each cycle comprising four parts: A) digging, B) raising and rotating the boom, C) unloading the soil, and D) returning. After the trenching task is completed, the operator pauses the machine, remains still with eyes closed for 1 minute, and then drives the machine 5 meters to the trench-finding exploration task area to begin the exploration task.

In the trench-finding task, a target working area of 1.5 meters by 2 meters is also designated in advance, and a steel pipe is pre-buried at a depth of 1.5 meters underground. The participant is required to use the excavator to perform trench-finding actions within the designated area. The trench exploration consists of six cycles, with each cycle comprising seven parts: A) extending the arm, B) lowering the boom, C) smoothly advancing the bucket, D) gradually closing the bucket, E) raising and rotating the boom, F) unloading the soil, and G) returning. During the trench-finding task, the participant must ensure that the pre-buried steel pipe is not touched throughout the six cycles. After completing the cycles, the participant must use the excavator to lift a steel pipe next to the experimental site, hold it for 3 seconds, and then return it to its original place. This action consists of six parts: A) lowering the boom, B) grabbing the steel pipe, C) raising the boom, D) maintaining level, E) lowering the boom, and F) placing the steel pipe. Both the trenching and trench-finding tasks are not restricted by time; the experiment is considered complete once all tasks have been performed.

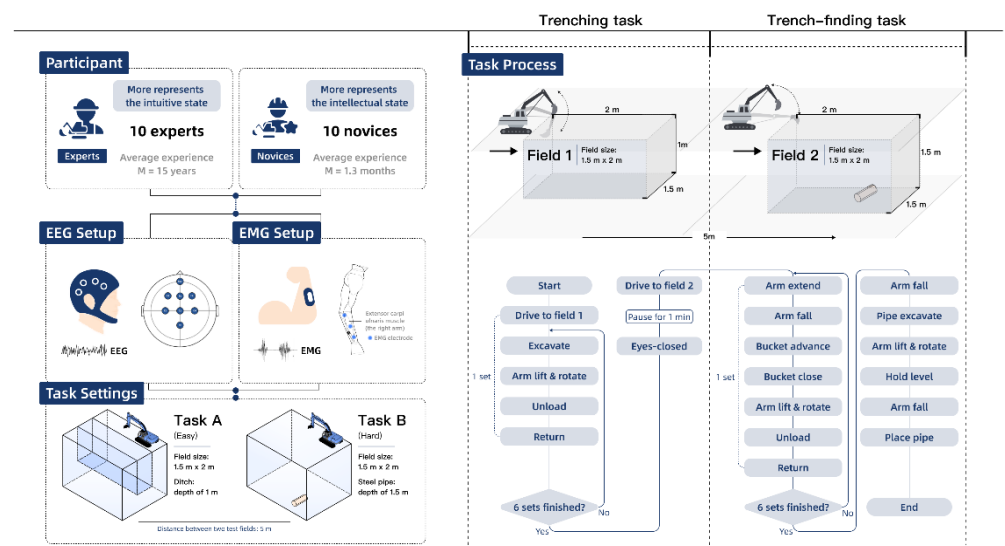

Figure 5 Experimental protocol

## 4.2 Participants

The experiment included 20 subjects, consisting of 10 expert machine operators with over 5 years of experience (average experience M = 15 years) and 10 novice machine operators with 1-6 months of experience (average experience M = 1.3 months). During the preliminary screening of the experimental data, it was found that a significant number of outliers appeared in the data of some subjects due to interference from mechanical vibrations, environmental noise, and other factors. As a result, these portions of the data was excluded from the analysis. The final dataset included 7 expert operators (work experience M = 13 years) and 7 novice operators (work experience M = 1.2 months) for data analysis.

The participants were all right-handed, with normal motor skills, normal or corrected-to-normal vision, and without color vision abnormalities, brain injury, or a history of neurological or psychiatric disorders. The study received an explicit ethics approval from the University Ethics Committee at the Tongji University. Additionally, we obtained informed consent from all participants as required by the Ministry of Education. All participants signed informed consent forms indicating their willingness to participate.

### 4.3 EMG and EEG setup

The ErgoLAB HME Synchronization Platform V3.0 was used to record participants' EEG and EMG signals. This system includes the Semi-dry Wearable EEG Measurement System, the ErgoLAB sEMG Wearable Surface Electromyography Device, and the ErgoLAB HME Synchronization Platform V3.0 software.

The Semi-dry Wearable EEG Measurement System was employed to record brain signals from participants. The electrode positions on the EEG cap were based on the International 10–20 system, with the ground electrode placed at the midpoint between the frontal and central regions, and the reference electrode at the intersection of the line connecting the ears and the sagittal line. EEG signals were recorded from eight electrode channels: Fpz, Fz, F3, F4, C3, Cz, C4, and Pz, with a sampling frequency of 1024 Hz and electrode impedance of 10 kΩ.

The ErgoLAB sEMG Wearable Surface Electromyography Device (Ergolab, Jinfa Technology, Beijing) was used to record the electromyographic activity of participants at a sampling rate of 1024 Hz. Conductive gel (Ag/AgCl, 3 cm in diameter) was applied to the motor point of the ulnar-side wrist extensor muscle of the dominant hand.

### 4.4 Data preprocessing
### 4.4.1 EEG preprocessing

EEG is widely employed as a performance indicator of cognitive behavior, serving as an external reflection of the brain's information processing activities [52]. Power Spectral Density (PSD) analysis can be used to calculate the power of specific frequency bands during the experiment and reflect different forms of cognitive processing and states of consciousness during the task execution of the subject [53]. Rhythmic alpha oscillations are typically associated with sensory, motor, and cognitive information processing [54,55]. Rhythmic beta oscillations are implicated in the regulation of attention within the human visual system and are associated with bodily activities [55]. Emotional stress, the management of working memory, and workload regulation are predominantly correlated with rhythmic theta oscillations [44]. Therefore, most previous studies have chosen PSD values of alpha, theta, and beta bands as the main indicators for analyzing and assessing subjects' cognitive load, attention management, and skill acquisition status. However, in the construction machinery industry, exemplified by a typical mechanical excavator, its low-frequency vibration range partially overlaps with the EEG theta band[56]. This overlap results in general abnormalities in the PSD values of the theta band, necessitating its exclusion from the experiment. In summary, the study utilized EEG spectrum analysis to calculate the power of specific frequency bands during the experiment, selecting the alpha and beta bands as typical indicators to further analyze the cognitive states and neural activity characteristics of both expert and novice groups during task execution.

In this study, the ErgoLAB system applied a 0.01~100 Hz high-low pass filtering and 60 Hz band-stop filtering to the collected EEG signals, with the EEGLAB plugin conducting independent component analysis to identify and eliminate non-brain artifacts like ocular, muscular, and cardiac interferences. Subsequently, the study utilized the EEG analysis module supported by the ErgoLAB system to generate PSD values for the alpha and beta bands, and averaged the derived PSD values for experts and novices. After comparing the PSD values of the two groups using the Mann-Whitney

U test, box plots were generated to illustrate the inter-group PSD values across different tasks. After normalizing the PSD values across frequency bands, the study used spherical spline interpolation to estimate the potentials in areas not directly covered by scalp electrodes, generating continuous scalp potential distribution maps. Brain topographic maps for expert groups and novice groups were created in the key alpha and beta frequency bands, with the intensity of electrical activity in different regions represented by color depth on the maps, visualizing the distribution and intensity differences of electrical activity.

### 4.4.2 EMG preprocessing

Corticomuscular Coherence (CMC) quantifies the coherence between cortical and muscular activities by integrating EEG or magnetoencephalography with EMG, and is the key to understanding how the brain regulates and controls bodily movements, providing insight into the degree of cortical involvement in motor control and the effectiveness with which movement execution integrates sensory information from the brain. [45]. In the acquisition and execution of motor skills, higher CMC generally indicates superior motor control abilities, suggesting more effective communication between the brain and muscles [57-59]. Neural oscillations during motor processes can primarily be observed in the beta band and gamma band. Beta-band CMC is associated with fine motor control [58,59], motor preparation [60], and sensorimotor integration [61]. Gamma-band CMC is related to proprioceptive feedback in more dynamic sensorimotor tasks [62,63] and the integration of cortical components during visuomotor paradigms [64,66]. In summary, CMC provides an effective measure of communication efficiency between the brain and muscles and can effectively reflect the perception-action coupling characteristics of the subjects. Therefore, this study selected CMC values in the beta and gamma bands as key indicators to evaluate the degree of brain-muscle coupling in novices and experts during the execution of tasks of varying difficulty, in order to assess their perception-action coupling characteristics.

The EEG signal sampling rate used in this study was 256 Hz, and the EMG signal sampling rate was 1024 Hz. The study utilized MATLAB to resample the EEG data exported from ErgoLAB to 256 Hz, ensuring that the signals could be compared on the same time scale. The study defined two frequency bands of interest: Alpha (8-12 Hz) and Beta (15-30 Hz). A sliding window length of 250 milliseconds was set to obtain sufficient frequency resolution (4 Hz), while also avoiding overlap between consecutive stimuli, with power spectra and coherence calculated at 20-millisecond increments. The data was divided into windows with a length equal to 1/8 of the total data length, with each window overlapping by half to balance time and frequency resolution. The study used the cpsd function in MATLAB's eeglab to calculate the cross-power spectral density between the EEG and EMG signals, and the pwelch function to calculate the power spectral density of the EEG and EMG signals.

The formula for calculating coherence is as follows:

$$C_{xy} = \frac{|P_{xy}|^2}{P_{xx}P_{yy}}$$

Here, Pxx and Pyy represent the power spectral density of the EEG and EMG signals, respectively, and Pxy is the cross-power spectral density. The coherence value ranges from 0 to 1, where 1 indicates complete synchronization and 0 indicates no synchronization.

The significance threshold for coherence at the 95% confidence level is calculated using the following formula:

$$C_{threshold} = \sqrt{1 - \alpha^{\frac{1}{df-1}}}$$

Here, α represents the significance level (1 - confidence level), and df is the degrees of freedom (approximately equal to the number of data segments). The significance threshold is used to determine whether the coherence is statistically significant. For each frequency band, the coherence frequency indices that exceed the significance threshold are identified. By calculating the area of coherence that exceeds the threshold within these bands, the degree of synchronization between the EEG and EMG signals in specific frequency bands can be quantified. The significant area within each band, which represents the integral of coherence values exceeding the threshold, is computed and recorded for all signals.

The coherence indices obtained were statistically analyzed using the Mann-Whitney U test. Additionally, box plots and topographic maps were generated to visually illustrate the data characteristics.

## 4.5 Experimental data results
### 4.5.1 Result of EEG assessment

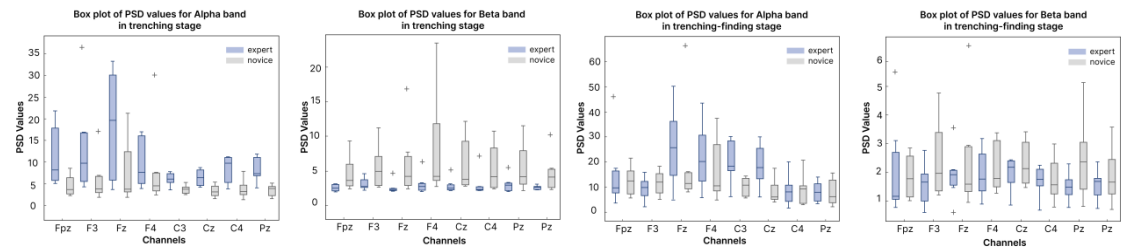

Figure 6　Box Plot of PSD Values for Different EEG Electrodes Between Experts and Novices in the Alpha and Beta Bands During the Trench-Finding and Trenching Stages

The results indicate that during the trench-finding task, experts exhibited significantly higher alpha band PSD values in the central, parietal, and prefrontal regions compared to novices, including at electrodes C3 (U = 33, p = 0.0111), Cz (U = 34, p = 0.0175), C4 (U = 33, p = 0.0111), and Fpz (U = 36, p = 0.0262) (as shown in Figure 6). Although differences in PSD values were observed at electrodes F3, Fz, and F4, the statistical analysis showed that these differences were not significant (p > 0.05). Despite the experts' median values generally being higher than those of the novices, the overall distribution was similar, with considerable overlap and comparable data ranges. The topographic map results show that experts had higher PSD values in the parietal and posterior parietal regions, especially near the parietal midline, whereas novices showed generally lower PSD values across the scalp (as shown in Figure 7). In the beta band, experts had significantly lower PSD values than novices at all electrodes except C4 (p > 0.05), including Fpz (U = 71, p = 0.0175), F3 (U = 70, p = 0.0262), Fz (U = 72, p = 0.0111), F4 (U = 71, p = 0.0174), C3 (U = 70, p = 0.0262), Cz (U = 70, p = 0.0262), and Pz (U = 71, p = 0.0175). The box plots indicate that novices displayed a wider interquartile range at these electrodes. Although some differences in PSD values were noted at the C4 electrode, the statistical analysis did not reach significance (p > 0.05). The topographic maps showed that overall beta activity was lower in experts during the trench-finding task, particularly in the central, prefrontal, and parietal regions (as shown in Figure 8). In contrast, novices exhibited more pronounced beta activity in the prefrontal and parietal areas.

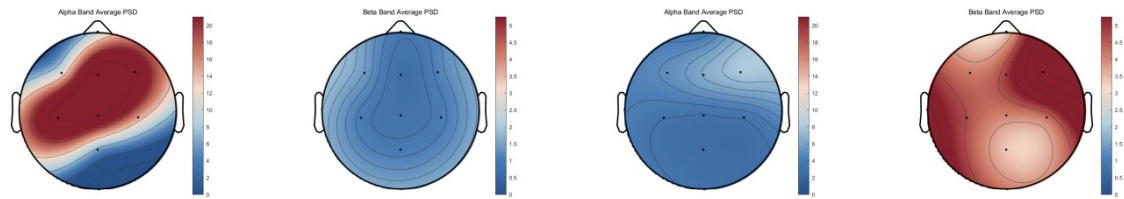

Figure 7 PSD Brain Topographic Maps for Experts and Novices in the Alpha and Beta Bands During the Trench-Finding Stage

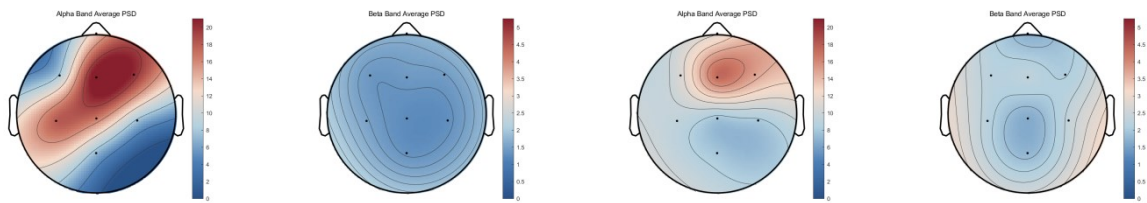

Figure 8 PSD Brain Topographic Maps for Experts and Novices in the Alpha and Beta Bands During the Trenching Stage

In the trenching task, within the alpha band, the PSD values for the expert group were generally higher, but these differences did not reach statistical significance (p > 0.05). In the beta band, the overall PSD values for experts were lower, with slightly higher values in the central region compared to peripheral areas; similarly, novices also had low PSD values, with slightly higher power density in the central and frontal regions, but these differences were also not statistically significant. There were no significant differences in EEG power spectral density distribution between experts and novices in the beta band (p > 0.05).

### 4.5.2 Result of CMC assessment

The results indicate that during the trench-finding task, experts exhibited higher CMC values in the beta band compared to novices, with statistically significant differences observed at F3 (U = 70, p = 0.0262), C3 (U = 71, p = 0.0175), and Cz (U = 71, p = 0.0175). Similarly, experts had higher CMC values in the gamma band than novices, with even more locations showing statistical significance, including Fpz (U = 70, p = 0.0262), C3 (U = 71, p = 0.0175), Cz (U = 71, p = 0.0175), C4 (U = 70, p = 0.0262), and Pz (U = 70, p = 0.0262). The box plots demonstrate that, in both the beta and gamma bands, experts had higher median CMC values and a wider distribution range across all channels and frequency bands, while novices showed relatively lower median coherence (as shown in Figure 9). Topographic maps revealed that, whether in the beta or gamma bands, experts' CMC values were significantly higher than those of novices, with the most pronounced differences observed in the left and posterior regions of the scalp. (as shown in Figure 11)

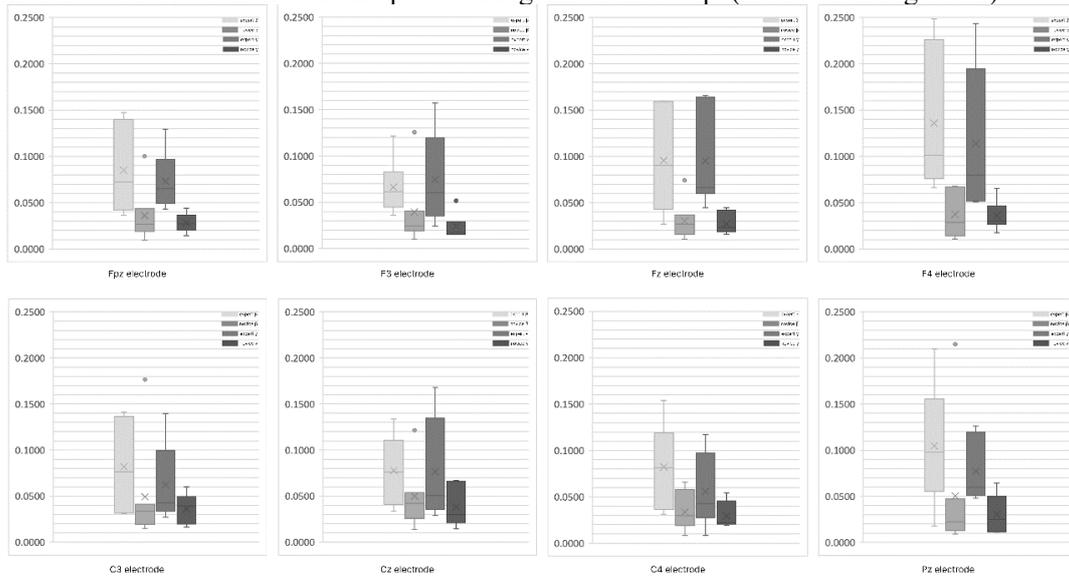

Figure 9 Comparison of Corticomuscular Coherence Across Different EEG Electrodes Between Experts and Novices in the Beta and Gamma Bands During the Trench-Finding Stage

In the trenching task, no significant differences were observed in CMC values between experts and novices in the beta and gamma bands (as shown in Figure 10)

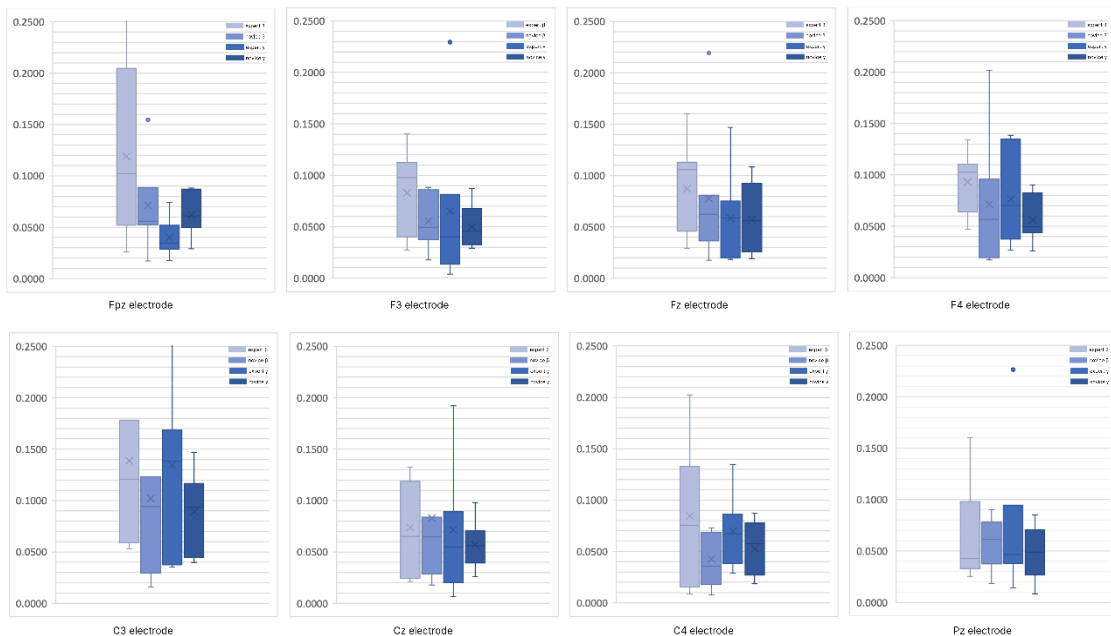

Figure 10 Comparison of Corticomuscular Coherence Across Different EEG Electrodes Between

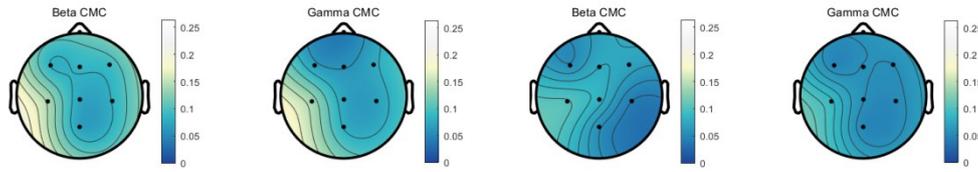

Figure 11 The Cortico-Muscular Topographic Map for Experts and Novices in the Beta and Gamma Bands During the Trench-Finding Stage

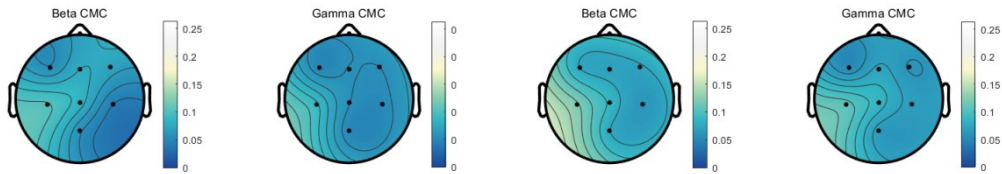

Figure 12 The Cortico-Muscular Topographic Map for Experts and Novices in the Beta and Gamma Bands During the Trenching Stage

## 4.6 Analysis
### 4.6.1 Differences in EEG characteristics between experts and novices
The experimental results indicate that during the trench-finding task, experts exhibited significantly higher PSD values in the alpha band in central and posterior regions, particularly in the central/parietal (C3, Cz, C4) and specific regions of the prefrontal cortex (Fpz). Alpha power in EEG is negatively correlated with cortical activation; an increase in alpha power reflects reduced brain activation, which is typically associated with lower working memory load and more efficient cognitive integration [66]. The higher PSD values in the alpha band among experts suggest their ability to suppress irrelevant distractions during the task, thereby improving the efficiency of cognitive resource utilization. In contrast, novices displayed generally lower PSD values in the alpha band, especially in the central and posterior regions. The lower activity in the alpha band may indicate that novices require more cognitive resources to understand and process complex information during the task, reflecting their weaker cognitive integration capabilities.

In the beta band, experts exhibited generally lower PSD values, particularly in the central regions (C3, Cz, C4) and prefrontal/frontal areas (Fpz, F3, Fz, F4). Increased beta activity is associated with heightened alertness and attention, indicating a high level of cognitive engagement, and is closely related to physical activity [67,68]. The lower beta activity observed in experts suggests that they required less alertness and selective attention during the task, reflecting a smoother task execution process with lower cognitive resource demands. In contrast, novices showed significantly higher PSD values in the beta band, especially in the prefrontal (Fpz, F3, Fz, F4) and parietal regions (C3, Cz, C4, Pz). The higher beta activity in novices indicates a greater need for alertness and selective attention during task execution, likely due to their lack of experience, requiring more cognitive resources to focus and process information effectively.

During the trenching task, in the alpha band, no significant differences were observed in PSD values between experts and novices in the central and parietal regions. This indicates that in a simple task, the level of cortical activation is similar between the two groups, suggesting that both experts and novices exhibit comparable levels of cognitive load. Similarly, in the beta band, there were no significant differences in PSD values between experts and novices in the central and prefrontal regions, indicating that the need for alertness and selective attention, as well as the cognitive resource demands for task execution, are largely consistent between the two groups in simple tasks.

### 4.6.2 Differences in CMC characteristics between experts and novices
CMC values reflect varying degrees of synchrony between the brain and muscles. Beta-band CMC

values are associated with fine motor control [57,58], motor preparation [59], and sensorimotor integration [60]. Higher beta-band CMC values indicate stricter cortical control over muscle activity and more effective integration between perception and action. The study found that experts exhibited significantly higher beta-band CMC values than novices at the central region electrode (Cz) and the left parietal region electrode (C3), particularly at the latter. The central region is closely related to motor control, and the high beta-band CMC in these channels suggests that experts are able to control limb movements more precisely, reflecting their precise control of motor units and effective transmission along the corticospinal tract, indicating more pronounced perception-action coupling characteristics. The C3 electrode is located in the motor cortex of the left hemisphere, responsible for controlling the right side of the body and is associated with fine motor control of the right hand [69]. The significant difference in this region may indicate that, in high-complexity tasks, experts exhibit more prominent perception-action coupling, effectively integrating external sensory feedback and adjusting muscle activity accordingly to achieve smooth and precise motor control. The lower beta-band CMC values in novices in the central region suggest a deficiency in motor control and sensorimotor integration, which may lead to imprecise motor control and lower task execution efficiency. The particularly low CMC values at the C3 electrode suggest that novices are relatively lack coordination in controlling right-hand movements.

In the gamma band, CMC values are typically associated with proprioceptive feedback during sensorimotor tasks [64,65] and the integration of cortical components during visuomotor paradigms [62,63]. High gamma-band CMC is usually linked to the rapid integration of visual, tactile, and proprioceptive information, indicating that experts can quickly react and adjust motor strategies [70]. In the gamma band, experts' high CMC in the prefrontal region reflects their efficient perception-action integration abilities when handling complex cognitive tasks. Experts' high gamma-band CMC in the central region electrode (Cz) and the left and right parietal region electrodes (C3, C4) suggests a high degree of neural coordination during tasks, enabling effective sensorimotor integration when executing complex motor tasks. In contrast, novices' low gamma-band CMC in the central region reflects their lack of effective neural coordination and motor control abilities. In complex tasks, novices often require more cognitive resources to focus on the task, leading to lower neural integration efficiency and difficulty maintaining a high level of perception-action coupling, preventing them from executing complex tasks as smoothly as experts.

In complex tasks, experts and novices exhibit significant differences in CMC values at certain electrodes in the beta and gamma bands; however, in relatively simple tasks like trenching, there are no significant differences in CMC values between experts and novices in both the beta and gamma bands. This suggests that in complex tasks, experts possess efficient cognitive control and precise motor control abilities during task execution. Extensive practice and repeated training have optimized the neural circuits and muscle control abilities of experts, allowing them to achieve higher task performance with fewer cognitive resources. This efficient perception-action coupling mechanism enables them to coordinate and execute complex tasks more precisely. The significantly lower CMC values in novices indicate their deficiencies in perception-motor integration, neural synchrony, and motor control. In simple tasks, where the demands are lower, both experts and novices rely on basic skills to complete the task without the need for significant cognitive resource investment, resulting in similar neural activity between the two groups.

### 4.6.3 Distinct physiological patterns between two states

Based on experimental observations and data analysis, there were no significant differences between experts and novices in cortical activation levels, cognitive load, selective attention, or perception-action coupling during the simple trenching task. This indicates that in simple tasks, the skill level required from the operator is relatively low, allowing both experts and novices to exhibit relatively smooth and automated behavior without noticeable differences in cognitive resource demand, neural coordination, sensorimotor integration, or motor control abilities. In contrast, during the trench-finding task, experts and novices displayed significant differences in operational and neural patterns. Experts maintained a highly fluent task execution state, requiring fewer cognitive resources and demonstrating stronger sensorimotor integration, while novices required significantly more cognitive resources and exhibited a reduction in overall task execution fluency.

In summary, during complex tasks, experts tend to exhibit lower cognitive load and higher perception-action coupling, corresponding to the hypothesized intuitive state. In contrast, novices demonstrate higher cognitive load and lower perception-action coupling, aligning with the hypothesized intellectual state. These two states can be further characterized by differences in EEG

PSD values and corticomuscular coherence (CMC) values, as outlined in Table 1.

Table 1: Characteristic Differences in Brain Electrical Power Spectral Density (PSD) and Corticomuscular Coherence (CMC) Values Corresponding to Intuitive and Intellectual States

| Human states | Physiological characteristics | | Band | Regions |
|---|---|---|---|---|
| State of intuition | Lower cognitive load and higher brain-muscle coupling | PSD | Higher alpha activity | Frontal and parietal lobes |
| | | | Lower beta activity | Frontal/prefrontal lobes, parietal lobes |
| | | CMC | Higher beta-band coherence | Central region, left parietal lobe |
| | | | Higher gamma-band coherence | Central region |
| State of intellect | Higher cognitive load and lower brain-muscle coupling | PSD | Lower alpha activity | Frontal and parietal lobes |
| | | | Higher beta activity | Frontal/prefrontal lobes, parietal lobes |
| | | CMC | Lower beta-band coherence | Central region, left parietal lobe |
| | | | Lower gamma-band coherence | Central region |

## 5 DISCUSSIONS

In this study, we addressed the practical challenges of HMC by integrating philosophical insights. We further defined the concept of intuitive interaction flow. Based on this concept, we developed a model for intelligent HMC. Centered around this concept and model, our research conducted preliminary experimental analyses to explore the underlying physiological correlates. Additionally, we proposed a potential pathway for the practical application of this model.

### 5.1 Concept definition and model contributions

The paradigm shift brought by Industry 4.0 has significantly impacted human-machine relationships. Early research on HMC primarily focused on traditional intelligent systems that could automatically respond to predefined scenarios, exploring how to better utilize their computational intelligence to assist humans [71,72].In these studies, intelligent systems primarily took over procedural tasks from humans. As machine automation and autonomy continue to advance, the complexity of HMC has increased. Machines are increasingly taking on tasks previously performed by humans, and HMC task allocation mechanisms have evolved beyond simply delegating functions and operations to machines [71]. Although machine automation offers numerous advantages, the unique cognitive framework for HMC still needs to consider human intelligence as an important component.

The in-depth analyses of human and machine intelligence by scholars such as Dreyfus have provided valuable insights into the distinct advantages of each and have prompted a rethinking of HMC. The human body plays a central role in human activity, and embodied skills cannot be easily replaced by mechanized processes. The fundamental distinction between human and machine intelligence lies in embodiment, where human-specific embodied skills and the intuitive states that emerge from skill acquisition are uniquely valuable in complex situations.

Building on these insights, this study further explored two typical behavior modes in HMI: skill-based intuitive behavior and knowledge-based intellectual behavior. The focus was on how to combine the 'intuitive' strengths of humans with the human-like 'intellectual' strengths of machines. This led to the innovative definition of the intuitive interaction flow concept and the development of a dual-loop HMC task allocation model. The proposed model aims to integrate machine intelligence into the process while preserving human self-efficacy, offering a new perspective on addressing HMC challenges in the context of Industry 4.0.

## 5.2 Physiological correlations and limitations of experimental analysis

Centered around the concept of intuitive interaction flow, this study sought to further analyze its physiological correlations. Previous research has identified human behavior and cognitive states as key parameters in the HMC loop, with indicators such as muscle activity, brain signals, and visual cues being used to help machine systems understand human states and achieve efficient, seamless collaboration [73-75]. Therefore, this study first analyzed the potential cognitive and behavioral characteristics corresponding to intuitive and intellectual states based on the LIDA model. Experts and novices were selected as typical experimental groups, and a comparative experiment was designed to summarize the physiological characteristics associated with these two typical states, supporting the identification and classification of human states within the proposed HMC model.

The study hypothesizes that differences between intuitive and intellectual states can be distinguished through EEG and EMG physiological indicators: experts are expected to exhibit more intuitive states during task execution, characterized by lower cognitive load and higher perception-action coupling, while novices are likely to exhibit more intellectual states, with higher cognitive load and lower perception-action coupling. Through comparative experiments and data analysis, the study preliminarily validated the hypothesis and provides an initial summary of the physiological indicators corresponding to the two states, laying the groundwork for further in-depth research and practical application of the model.

However, a major limitation of this study is the small number of participants, and the potential interference from noise and vibration in real operational scenarios with heavy machinery was not fully excluded. The analysis in this experiment only included data from 14 participants. Future work should involve replicating the study with a larger dataset to confirm these findings.

## 5.3 Limitations and future work

Machine learning algorithms such as random forests, support vector machines (SVM), and various neural networks have been proven effective and feasible in predicting and understanding human behavior and cognitive states in human-machine interactions. In HMC models, detecting, recognizing, and classifying the cognitive and behavioral states of human subjects enables the implementation of adaptive machine collaboration and assistance strategies, thereby enhancing the flexibility and adaptability of HMC. Based on this, the study proposes an adaptive framework for intelligent systems from a collaborative perspective, indicating possible pathways for the practical application of this model.

However, the proposed intelligent system's adaptive framework requires further development for practical application. Further research opportunities include the continuous collection and processing of physiological data from heavy machinery operators to demonstrate the applicability and reliability of EEG and EMG measurements. Additionally, selecting appropriate machine learning algorithms based on the characteristics of the actual data and identifying the most suitable EEG and EMG indicators through a thorough evaluation of these algorithms will be essential. In the future, research could be further refined to more thoroughly explore the application of neurophysiological data and machine learning techniques in optimizing HMC systems, as well as to comprehensively validate the proposed HMC model over the long term.

## 6 CONCLUSIONS

The advancements in AI technologies have created new opportunities for Industry 4.0. This study integrated philosophical theories and cognitive science perspectives to define the concept of intuitive interaction flow and constructed an innovative HMC model. This model incorporates a dual-loop dynamic of intuition and intellect, highlighting the value of combining human intuition with machine intelligence to address complex tasks in an automated and intelligent context.

Based on the experimental results, this study summarized the cognitive and behavioral state differences, as well as EEG and EMG activity characteristics, between experts and novices performing the same tasks, providing preliminary validation of the research hypotheses. Future research should refine the methods and design, empirically classify the neural activity characteristics corresponding to the two behavioral modes, and validate the model's effectiveness. This will provide scientific evidence for optimizing task allocation strategies in HMC. By combining theoretical deduction and empirical analysis, this study emphasizes the importance of effectively integrating human intuition with machine intelligence processing capabilities in HMC systems, offering a new perspective on HMC.

## ACKNOWLEDGMENTS
The authors want to thank all the participants to the study for their availability.

## FUNDING
This work was supported by the National Key Research and Development Program of China [Grant Number 2021YFF0900602] and The 2 Batch of 2023 MOE of PRC Industry-University Collaborative Education Program [Program No. 220705329315952, Kingfar-CES "Human Factors and Ergonomics" Program]